\documentclass{article}
\pdfoutput=1
\usepackage{amssymb}
\usepackage{amsbsy}
\usepackage{enumerate}
\usepackage{graphicx}
\usepackage{amsthm}
\usepackage[T1]{fontenc}
\usepackage[latin9]{inputenc}
\usepackage{amsmath}
\usepackage{mathabx}
\usepackage{color}
\usepackage{hyperref}
\usepackage{booktabs}
\usepackage{comment}

\usepackage{cancel}

\newcommand{\be}{\begin{equation}}
\newcommand{\ee}{\end{equation}}

\def\bea{\begin{eqnarray}}
\def\eea{\end{eqnarray}}
\def\bean{\begin{eqnarray*}}
\def\eean{\end{eqnarray*}}

\def\l2{\log_2\,}
\newcommand{\barr}{\begin{array}}
\newcommand{\earr}{\end{array}}

\newcommand{\bed}{\begin{displaymath}}
\newcommand{\eed}{\end{displaymath}}
\newcommand{\bal}{\begin{array}{ll}}
\newcommand{\eal}{\end{array}}

\def\mc#1{\mathcal#1}

\makeatother

\usepackage[square, comma, numbers, sort&compress, merge, mcite]{natbib}
\usepackage{mciteplus}
\mciteSetSublistMode{n}

\begin{document}
\date\time

\title{\hfill ~\\[-30mm]
       \hfill\mbox{\small }\\[30mm]
       \textbf{ An Asymmetric TBM Texture}
       } 
\date{}
\author{\\ Moinul Hossain Rahat,\footnote{E-mail: {\tt mrahat@ufl.edu}}~~Pierre Ramond,\footnote{E-mail: {\tt ramond@phys.ufl.edu}}~~Bin Xu,\footnote{E-mail: {\tt binxu@ufl.edu}}\\ \\
  \emph{\small{}Institute for Fundamental Theory, Department of Physics,}\\
  \emph{\small University of Florida, Gainesville, FL 32611, USA}
  }

\maketitle

\begin{abstract}
\noindent   
We construct a texture where the Seesaw matrix is diagonalized by the TriBiMaximal (TBM) matrix with a phase. All CKM and PMNS angles are within their pdg values, and the mass relations of quarks and charged leptons extrapolated to GUT scale are satisfied, including the Gatto relation. The novel ingredient is the asymmetry of the down-quark and charged lepton Yukawa matrices. Explaining the reactor angle requires a \cancel {CP} phase in the TBM matrix, resulting in the Jarlskog-Greenberg invariant at $|J|=0.028$, albeit with an undetermined sign. While $SO(10)$ restrains the right-handed neutrino Majorana matrix, the neutrino masses  are left undetermined. 

\end{abstract} 

\thispagestyle{empty}
\vfill
\newpage
\setcounter{page}{1}
\section{Introduction}
The Standard Model evocates simplicity at smaller scale,  both in gauge couplings (grand-unified theories), and in the mass patterns of down-quarks and charged leptons. Yet the mixings of leptons and quarks are starkly different: neutrino oscillations \cite{davis1968search,*fukuda,*ahmad,*eguchi,*Mohapatra,*Lavignac} require two large lepton mixing angles.  

Quark-lepton mixing disparity, anticipated in  the $SO(10)$-inspired seesaw mechanism \cite{minkowski1977,*Gell-Mann_Ramond_Slansky,*yanagida,*glashow1979cargese}, yields tiny neutrino masses through the ratio of electroweak to grand-unified scale.

A pretty matrix (with an ugly name Tri-Bi-Maximal (TBM)) \cite{harrison2002tri, *tbm2},   

\be
\mc U^{}_{\rm TBM}=\begin{pmatrix}\sqrt{\dfrac{2}{3}}&\dfrac{1}{\sqrt{3}}&0\\[0.8em] -\dfrac{1}{\sqrt{6}}&\dfrac{1}{\sqrt{3}}&\dfrac{1}{\sqrt{2}}\\[1em] 
\dfrac{1}{\sqrt{6}}&-\dfrac{1}{\sqrt{3}}&\dfrac{1}{\sqrt{2}}\end{pmatrix}
\ee
diagonalizes the seesaw matrix. Its two large angles near their pdg values suggest a discrete crystalline flavor symmetry   at  GUT scale (see \cite{altarelli2010discrete, *king2013neutrino, *tanimoto2015neutrinos, *meloni2017gut, *petcov2017discrete} for recent reviews).  TBM simplicity comes at a cost: the reactor angle $\theta_{13}$ generated from the seesaw is zero.

The PMNS lepton mixing matrix is an overlap of  seesaw  and charged lepton mixing matrices. The  latter, derived from the charged lepton Yukawa matrix,  may generate enough  ``Cabibbo Haze'' \cite{cabibbohaze} to explain the data, but symmetric Yukawa matrices  with TBM diagonalization underestimate the reactor angle \cite{kile}. Authors who assume symmetric Yukawa matrices require seesaw diagonalization beyond TBM \cite{harrison2014deviations, *ross2018}.

\vskip .5cm
In this paper we argue for seesaw simplicity of TBM diagonalization by seeking textures where the value of the reactor angle \cite{doublechooz, *dayabay,  *reno} is fully explained by Cabibbo haze.  

Our objective is to search for asymmetric Yukawa matrices which satisfy all experimental constraints: the CKM matrix, the Gatto relation \cite{gatto} $\tan{\theta_C}=\sqrt{m_d/m_s} \approx \lambda$ where $\theta_C$ is the Cabibbo angle and $\lambda$ is a Wolfenstein parameter, the down-quark and charged lepton mass relations at GUT scale using the renormalization group \cite{chanowitz1977price} $m_b=m_\tau$, $m_d=3m_e$, $m_s=m_\mu /3$
together with the values of the three lepton mixing angles \cite{pdglive, *pdg}. Our bottom-up approach makes extensive use of the patterns suggested by the $SU(5)$ and $SO(10)$ grand-unified groups. 

The main result of this paper is the construction of a specific $3\times 3$ Yukawa matrix that has a simple asymmetry of $\mathcal{O}(\lambda)$ in the $(31)$ matrix element. The Cabibbo haze generated by this texture, together with TBM seesaw diagonalization, leads us to slight overestimation of the reactor angle. However, the introduction of a single \cancel{CP} phase in the TBM matrix fits the reactor angle to data, while simultaneously bringing the atmospheric and solar angles within their pdg bound. This nontrivial TBM phase generates \cancel{CP} phase $\delta_{CP} = \pm 1.32 \pi$ in the PMNS matrix, resulting in Jarlskog-Greenberg invariant \cite{jarlskog1, *greenberg1} $|J|= 0.028$. The sign comes from the seesaw; it is not specified by our texture.

\vskip 0.5cm
After a review of the salient Yukawa patterns suggested by $SU(5)$ and $SO(10)$, we discuss  a symmetric Georgi-Jarlskog texture to motivate our procedure. We then construct our asymmetric texture. A discussion of its implications for further theoretical construction follows. The uniqueness of the asymmetry in explaining the data are extensively discussed in the appendices.

\section{The Electroweak Sector} \label{sec2}
Quarks and charged lepton masses and mixings stem from the Standard Model's Yukawa matrices. 
We use the basis where the up-quark matrix is diagonal: $
Y^{(2/3)}_{}\sim m_t \,{\rm diag}(\epsilon^4,\,\epsilon^2,\,1)
$,  with $\epsilon \approx \lambda^2$ implying the large top quark mass. The down-quark and charged lepton Yukawa matrices are diagonalized by 
\be
Y^{(-1/3)}_{}=\mc U^{(-1/3)}_{}\mc D^{(-1/3)}_{}\mc V^{(-1/3)\dagger}_{},\quad 
Y^{(-1)}_{}=\mc U^{(-1)}_{}\mc D^{(-1)}_{}\mc V^{(-1)\dagger}_{}.
\ee
{\color{black} where $\mc U^{(q)}$ and $\mc V^{(q)}$ are unitary matrices. In this basis $\mc U^{(-1/3)} = \mc U^{}_{\mathrm{CKM}}$.}
{\color{black} At GUT scale ($10^{15}~\mathrm{GeV}$), renormalization group running yields simple diagonal mass matrices of the form
$$ {\mc D^{(-1/3)}}\sim m^{}_b\begin{pmatrix}\lambda^4/3&0&0\\0&\lambda^2/3&0\\0&0&1\end{pmatrix},\quad {\mc D^{(-1)}}\sim m^{}_\tau\begin{pmatrix}\lambda^4/9&0&0\\0&\lambda^2&0\\0&0&1\end{pmatrix},$$
up to  signs; all entries are expressed in terms of $\lambda$, the tangent of the Cabibbo angle $\theta_C$. A direct consequence of grand unification is $m_b=m_\tau$ at the GUT scale.  {\color{black} The  Gatto relation  linking a mixing to a ratio of eigenvalues is explicit from the above mass matrices.} Also from $ m_e\ m_\mu\approx m_d \ m_s$, we notice that  

\begin{align}
    \det Y^{(-1/3)}_{}\approx\det Y^{(-1)}_{} \label{detcond}
\end{align}
}
In $SU(5)$, the particles of each family are assigned to
$ {\bf\bar 5}=[\,{\bar d},(\nu_e,e)\,],~ {\bf 10}=[\,({u},{d}), {\bar u}, \bar e\,]$
so that the up-quark masses reside in $ {\bf 10}\cdot{\bf 10}={\bf \bar 5}_s+{\bf \overline{50}_s}+{\bf \overline{45}}_a$
while  the charged lepton and down-quark masses are in $ {\bf\bar 5}\cdot{\bf 10}={\bf 5}+{\bf 45}.$

There are  four Yukawa matrices $Y^{\bf \bar 5}_{},\, Y^{\bf 5}_{},\, Y^{\bf 45},\,Y^{\bf \overline {45}}$, 
so that $Y^{(-1/3)}$ and $Y^{(-1)}$ are related to ~$Y^{\bf 45}$ and $Y^{\bf 5}$ {\color{black}as 
\begin{align}
Y^{(-1/3)} = Y^{\bf 5} + Y^{\bf 45}, \qquad Y^{(-1)} = {Y^{\bf 5}}^T - 3{Y^{\bf 45}}^T 
\end{align}}
A simple combination of vacuum values due to Georgi and Jarlskog \cite{gj} yields the Gatto relation and 

\be
\label{gatto}
m_b=m_\tau,\,\, \dfrac{m_d}{m_s}=9\dfrac{m_e}{m_\mu}, \,\, m_\mu=3m_s,
\ee
all at Grand-Unified scale.
\vskip .5cm
In ${SO(10)}$ a right-handed neutrino $\widebar{N}$ is appended to each family, fitting in its spinor representation ${\bf 16}={\bf\bar 5}+{\bf 10}+{\bf 1}$.   Masses are generated by three couplings, since  ${\bf 16}\cdot{\bf 16}={\bf 10_s}+{\bf 126_s}+{\bf 120_a}$,
and three Yukawa matrices $Y^{\bf 10}_{},\quad Y^{\bf 126}_{},\quad Y^{\bf 120}_{}$. The new features are

\noindent - a  $(\Delta I_{\rm w}=0)$  {Majorana} mass matrix ${\mc M}$ with couplings  ${\mc M}\cdot \widebar{N} \widebar{N}$

\noindent - a  $(\Delta I_{\rm w}=\dfrac{1}{2})$ Yukawa  matrix $Y^{(0)}$ for neutrino Dirac masses. {\color{black} Minimal models of $SO(10)$ predict $Y^{(2/3)}_{}~\sim~Y^{(0)}_{}$.}
\vskip 0.3cm
The resulting mass structures are summarized in the following table
\begin{table}[ht]\centering
\caption{\color{black}Masses from couplings}
\vskip .2cm
\renewcommand{\arraystretch}{2}
\hspace{0cm}\begin{tabular}{@{}l l l l l @{}} 
\toprule
{ Masses}&{\!\!\!\! $SO(10)\! \supset\!  SU(5) \!\times\! U(1)$\!\!\!\!} &{\hspace{-0.2cm} $\bf{10}$}& $\bf{126}$&{${\bf{120}}$}\\
\midrule

{ Majorana Singlet ${\mathcal M}$}&{${\bf 1}_{-5}\cdot{\bf 1}_{-5}$}&-&{${\bf 1}_{10}$ }& -\\

{ Majorana Triplet}&{${\bf \bar 5}_3\cdot{\bf \bar 5}_3$}&-& {${\bf 15}_{-6}$ } &-\\

{ Dirac $\nu$ mass}&{${\bf 1}_{-5}\cdot{\bf \bar 5}_{3}$}&{${\bf  5}_{2}$}&  - &{${\bf  5}_{2}$}\\

{ up-quark}&{${\bf 10}_{-1}\cdot{\bf 10}_{-1}$}&{${\bf  5}_{2}$}&  -&{${\bf  45}_{2}$}\\

{ down-quark \& charged lepton}&{${\bf \bar 5}_{3}\cdot{\bf 10}_{-1}$}&{${\bf  \bar 5}_{-2}$}&{  ${\bf  \overline {45}}_{-2}$} &{${\bf  \overline {45}}_{-2}$}\\

\bottomrule
\end{tabular}
\end{table}

{\color{black} The second column shows fermion-fermion couplings in $SU(5)$ language, with the subscripts denoting $U(1)$ quantum numbers. The next three columns show possible Brout-Englert-Higgs (BEH) boson quantum numbers, coming from $\mathbf{10}$, $\mathbf{126}$ or $\mathbf{120}$ of $SO(10)$. For example, down-quark and charged lepton masses are generated by coupling ${\bf \bar 5}_{3}\cdot{\bf 10}_{-1}$ fermions to either a ${\bf  \bar 5}_{-2}$ (contained in $\bf 10$ of $SO(10)$) or a ${\bf  \overline {45}_{-2}}$ (contained in $\bf 120$ or $\bf 126$ of $SO(10)$) BEH boson.}
\vskip .5cm
{\color{black} Having done a general analysis of the electroweak input to the flavor jungle, we now proceed to the brief discussion of a symmetric texture that shows its inconsistency  with TBM mixing.}

\section{A Generic Georgi-Jarlskog Symmetric Texture}\label{GJ sym}
{\color{black}Our analysis of textures will follow a bottom-up approach that relies heavily on the grand-unified structures evocated by the Standard Model}. All parameters are expressed \`a la Wolfenstein in terms of the Cabibbo angle $\lambda$. 

In $SU(5)$ thre are two types  of Yukawa couplings, $\bf 5$ and $\bf 45$. In appendix \ref{appA}, we derive their  form

\be 
Y^{\bf 5}_{}=\begin{pmatrix}0&a\lambda^3&b\lambda^3\\
a\lambda^3&0&g\lambda^2\\
b\lambda^3&g\lambda^2&1\end{pmatrix},\qquad 
Y^{\bf 45}_{}=\begin{pmatrix}0&0&0\\ 0&c\lambda^2&0\\
0&0&0\end{pmatrix},\ee
where the prefactors $a, b, c, g \sim \mathcal{O}(1)$. The down-quark and charged lepton couplings follow,

\be
Y^{(-1/3)}_{}=\begin{pmatrix}
0&a\lambda^3&b\lambda^3\\
a\lambda^3&c\lambda^2&g\lambda^2\\
b\lambda^3&g\lambda^2&1
\end{pmatrix},\quad 
Y_{}^{(-1)}=\begin{pmatrix}
0&a\lambda^3&b\lambda^3\\
a\lambda^3&-{ 3}c\lambda^2&g\lambda^2\\
b\lambda^3&g\lambda^2&1
\end{pmatrix}
\ee

\noindent The prefactors (neglecting for now the \cancel{CP} phase) are identified with the Wolfenstein parameters \cite{wolfenstein}, 

\be 
a=\dfrac{1}{3},\,  c=\dfrac{1}{3},\,  b={A}\sqrt{\rho^2+\eta^2}=  0.306,\,g=A= 0.811
\ee
in such a way as to reproduce the  CKM matrix,  the Gatto relation and the  GUT scale mass ratios Eq.(\ref{gatto}). {\color{black}A systematic way to calculate the prefactors has been discussed in detail in appendix \ref{appA}.}

Since the Yukawa matrices are symmetric, the mixing matrix of the left-handed charged leptons is closely related to the CKM matrix, 

\be
\label{symmetric}
\mc U^{(-1)}_{}=\mc U_\mathrm{CKM}(c\rightarrow -3c),
\ee
according to $SU(5)$. The lepton mixing angles of the PMNS matrix are now extracted, assuming TBM seesaw diagonalization

\be
\label{PMNS}
\mc U^{}_{\mathrm{PMNS}}=\mc U^{(-1)T}_{}{\mc U^{}_{\mathrm{TBM}}}, \ee
where, neglecting the \cancel{CP} phase,

\be
 \mc{U}^{}_{\mathrm {PMNS}}=\begin{pmatrix}c^{}_{12}c^{}_{13}& 
s^{}_{12}c^{}_{13}&s^{}_{13} \\
-s^{}_{12}c^{}_{23}-c^{}_{12}s^{}_{23}s^{}_{13}&
c^{}_{12}c^{}_{23}-s^{}_{12}s^{}_{23}s^{}_{13}&s^{}_{23}c^{}_{13}\\
s^{}_{12}s^{}_{23}-c^{}_{12}c^{}_{23}s^{}_{13}&-c^{}_{12}s^{}_{23}-s^{}_{12}c^{}_{23}s^{}_{13}&c^{}_{23}c^{}_{13}\end{pmatrix}
\ee
Here $c_{ij} \equiv \cos{\theta_{ij}}$ and $s_{ij} \equiv \sin{\theta_{ij}}$.
Then, Eq.(\ref{symmetric}) and \eqref{PMNS} yields  

\bea
\label{reactor}
|\sin\theta^{}_{13}|&=&\dfrac{1}{\sqrt 2}\left|\mc{U}_{21}^{(-1)}+\mc{U}_{31}^{(-1)}\right|\\
&=&\dfrac{\lambda}{3\sqrt 2}(1-A\lambda^2)\approx \dfrac{\lambda}{3\sqrt 2}=0.051,
\eea
one third of its  pdg value $0.145$\footnote{The addition of the CKM phase will give an $\mathcal{O}(\lambda^5)$ correction.}. Symmetric Yukawa matrices and TBM seesaw diagonalization are
incompatible with data. 

\section{Asymmetric Textures}\label{sec3}
Seesaw TBM diagonalization requires asymmetric couplings in the input Yukawa matrices {\color{black}to be compatible with neutrino mixing angle data}. Eq.(\ref{reactor}) indicates that  TBM  with a larger reactor angle demands larger $\mc{U}_{21}^{(-1)}$ and/or  $ \mc{U}_{31}^{(-1)}$. Furthermore,   $\mc{U}_{21}^{(-1)}$  describes mixing between the two lightest families that is already large, so that increasing  $ \mc{U}_{31}^{(-1)}$  is most likely to yield the desired effect in $\theta_{13}$. 

The link between $\mc U^{(-1)}$ and  the CKM matrix of Eq.(\ref{symmetric}) must be loosened. This readily occurs for asymmetric matrices, 
\be
\mc{U}^{(-1)}=\mc{V}^{(-1/3)}(c\rightarrow-3c),
\ee
with  unknown $\mc V^{(-1/3)}$. 

\noindent The  asymmetry may be in the $\bf 45$ and/or $\bf 5$ couplings. 
\vskip .3cm
\noindent - The analysis of appendix \ref{appA} indicates that the $\bf{45}$ coupling in $(22)$ position of  $Y^{\bf 45}$ leads us to the correct mass ratios and CKM angles at GUT scale, as in the Georgi-Jarlskog construction. $\bf 45$ couplings in different places fail in one way or another, in particular for symmetric or antisymmetric off-diagonal couplings.  
\vskip .3cm
\noindent - {\color{black}With off-diagonal $\mathbf{45}$ couplings ruled out,} the asymmetry must be in the $\bf 5$ couplings. 
\vskip .3cm
\noindent Asymmetries split into three generic cases, along the $(12)-(21)$, $(23)-(32)$ and $(13)-(31)$ axes. Assume for simplicity that it appears in only one. In appendix \ref{appB} we show that an asymmetry along $(12)-(21)$ or $(23)-(32)$ does not alleviate the $\theta_{13}$ deficiency. The asymmetry must then reside in the $(13)-(31)$ axis of the $\bf 5$ couplings. 

To make it as large as possible, we  insert a term of $\mc O(\lambda)$ in the $31$ position,

\be 
Y^{\bf 5}_{}=\begin{pmatrix}bd\lambda^4&a\lambda^3&b\lambda^3\\
		a\lambda^3&0&g\lambda^2\cr
		d\lambda&g\lambda^2&1\end{pmatrix},\qquad 
Y^{\bf 45}_{}=\begin{pmatrix}0&0&0\cr 0&c\lambda^2&0\\
		0&0&0\end{pmatrix}.\ee
{\color{black}where now $a, b, c, d,$ and $g$ are $\mathcal{O}(1)$ prefactors. These are the input parameters of our texture.} Note that the $11$ term is explicitly inserted as it is of lower order than in the symmetric case, and to make the determinant with cofactor in the $22$ position vanish. The Yukawa  determinant equality Eq.(\ref{detcond}) is now satisfied.
\vskip 0.2cm
\noindent The Yukawa matrices of the down-quarks and charged leptons follow,
\be
Y^{(-1/3)}_{}=\begin{pmatrix}
		bd\lambda^4&a\lambda^3&b\lambda^3\\
		a\lambda^3&c\lambda^2&g\lambda^2\\
		d\lambda&g\lambda^2&1
\end{pmatrix},\qquad 
Y_{}^{(-1)}=\begin{pmatrix}
		bd\lambda^4&a\lambda^3&d\lambda\\
		a\lambda^3&-{ 3}c\lambda^2&g\lambda^2\\
		b\lambda^3&g\lambda^2&1
\end{pmatrix}
\ee

\noindent The prefactors are expressed in terms of the Wolfenstein parameters so as to reproduce the  CKM matrix, the  GUT scale mass ratios, and the Gatto relation, 

\be\label{prefactors}
a=\dfrac{1}{3},\,\,c=\dfrac{1}{3},\,\, g=A,\,\, b=A\sqrt{\rho^2+\eta^2},\,\, d=\dfrac{2a}{g}=\dfrac{2}{3A}.
\ee
\vskip .5cm
The new charged lepton mixing matrix, 

\be
\mc U^{(-1)}_{}=\begin{pmatrix}
1-\left(\dfrac{1}{18}+\dfrac{2}{9 A^2}\right) \lambda^2 & \dfrac{1 }{3} \lambda & \dfrac{2  }{3 A}\lambda\cr
- \dfrac{1 }{3} \lambda & 1-\dfrac{1}{18}\lambda ^2 & A \lambda ^2\cr
-\dfrac{2  }{3 A}\lambda & -\left(A+\dfrac{2}{9 A}\right) \lambda^2 & 1-\dfrac{2 }{9 A^2}\lambda ^2
\end{pmatrix}+\mc{O}(\lambda^3),
\ee
has extra elements of $\mc O(\lambda)$, which bring the reactor angle to a new value 

\be 
\sin \theta_{13}=\frac{\lambda}{3\sqrt{2}}\left(1+\frac{2}{A}\right)=0.184,
\ee
that is above its  pdg value by $2.26^\circ$\footnote{An  asymmetry of  $\mc O(\lambda^2)$  leaves the reactor angle well below its pdg value.}. 

The other two lepton mixing angles are also off their pdg values,
\be
\label{nophaseangles}
\theta_{12}=39.81^\circ \ (6.16^\circ\text{ above pdg}),\qquad 
\theta_{23}=42.67^\circ \ (2.90^\circ\text{ below pdg})
\ee
\vskip .5cm
The distinguishing feature of this asymmetry is a reactor angle above its experimental value. The addition of a \cancel{CP} phase \cite{chen2014cp} in the TBM matrix can be used to  lower \cite{petcov2015predicting} $\theta_{13}$ to its pdg value. 
\vskip .5cm
What makes this particular texture noteworthy is that by lowering the reactor angle to its experimental value, we not only find an   amount of CP-violation that is consistent with experiment, but also align both solar and atmospheric angles to their pdg values.
 
 We do not need to include the Majorana phases \cite{schechter1981neutrino} which enter only in total lepton-number violating physics. Of the many ways to insert phases in the TBM matrix, we choose 

\be
\label{TBM}
{\mc U}^{}_{\mathrm{TBM}}(\delta_{})=\begin{pmatrix}\sqrt{\dfrac{2}{3}}&\dfrac{1}{\sqrt{3}}&0\cr \\[-0.8em] -\dfrac{1}{\sqrt{6}}&\dfrac{1}{\sqrt{3}}&
\dfrac{ 1}{\sqrt{2}}\cr \\[-0.8em]
\dfrac{ {e^{i\delta_{}}}}{\sqrt{6}}&-\dfrac{ {e^{i\delta_{}}}}{\sqrt{3}}&\dfrac{e^{i\delta_{}}}{\sqrt{2}}\end{pmatrix}.
\ee
Neglecting the CKM phase,

\be
\mc{U}^{}_{\mathrm{PMNS}}=\mc{U}^{(-1)T}_{} \,\mc{U}^{}_{\mathrm{TBM}}(\delta)
\ee

\noindent The value of $\theta_{13}$ is lowered by the TBM phase\footnote{The CKM phase gives an $\mathcal{O}(\lambda^3)$ contribution, which is too small to affect the result} to,

\be
|\sin \theta_{13}|=\dfrac{1}{\sqrt 2}\left|\mc{U}_{21}^{(-1)}+\mc{U}_{31}^{(-1)}e^{i\delta_{}}\right| 
\leq \dfrac{1}{\sqrt 2}\left(\left|\mathcal{U}_{21}^{(-1)}\right|+\left|\mc{U}_{31}^{(-1)}\right|\right),
\ee
or, in terms of the Wolfenstein parameters, 

\be
\label{phasewolf}
|\sin \theta_{13}|=\frac{\lambda}{3\sqrt 2}\left|1+\frac{2e^{i\delta_{}}}{3 A}\right|+\mathcal{O}(\lambda^3).
\ee
We fit  $\theta_{13}$ to its central pdg value by using Eq.(\ref{phasewolf}), and find

\be  
\cos \delta_{}\approx 0.2,~~~\delta_{}=\pm 78^\circ.
\ee
The sign is undetermined at this stage.

A straightforward computation yields the remaining PMNS angles,
\be 
\label{anglewithphase}
\theta_{12} = 34.16^\circ~~(0.51^\circ~{\rm above ~pdg}), \qquad \theta_{23} = 44.91^\circ~~ (0.66^\circ~{\rm below ~pdg})
\ee 
to be compared with Eq.(\ref{nophaseangles}). 

{\color{black} The phase in TBM is carried to the PMNS matrix and generates the \cancel{CP} phase 
\be
\delta_{CP} = \pm 1.32\pi
\ee
This leads to the Jarlskog-Greenberg invariant
\be 
|J| = 0.028
\ee
Both $\delta_{CP}$ and $|J|$ are consistent with the current pdg value. It should be noted that the sign of $\delta_{CP}$ and $J$ cannot be determined from our texture, but rather by the hitherto unknown physics of the seesaw sector.}

A numerical summary of the texture can be found in appendix \ref{appC}.

\section{Theoretical Outlook}
The asymmetric TBM texture we just constructed provides an experimentally successful link between the electroweak Yukawa matrices and the seesaw scale Majorana mass matrix of the right-handed neutrinos. Both structures present  new theoretical patterns which we briefly address below. 

\vskip .5cm
\noindent {\large Yukawa Couplings}
\vskip .5cm

\noindent The crucial ingredient  is an asymmetric  $\mc O(\lambda)$ term in the $31$ element of the $SU(5)$ quintet Yukawa matrix $Y^{\bf \bar 5}$. 

It can arise from the vacuum value of one BEH boson $H^{\bf\bar 5}$, with  the symmetric and antisymmetric couplings canceling  (adding) in the $13$ ($31$) position. However, this is not technically natural in the absence of further symmetries. 

One simple remedy is to introduce two BEH bosons $H^{\bf\bar 5}$ and $H'{}^{\bf\bar 5}$, with a $Z_2$ exchange symmetry $
H^{\bf\bar 5}\,\longleftrightarrow\,H'{}^{\bf\bar 5}$. This insures equality between the symmetric and antisymmetric couplings. The desired cancellation occurs when the two vacuum values respect the $Z_2$ symmetry. 

The next step is to single out the (13)-(31) axis in the Yukawa matrix. One can simply add only this specific coupling to the Lagrangian, or seek a symmetry-based explanation which points those BEH bosons in the right flavor direction. 

A possible understanding appears natural with a $T_7$  discrete symmetry \cite{t7luhn}: {\color{black}the three families form a $T_7$ triplet, thus the Kronecker product of two fermions yield anti-triplet of $T_7$ in off-diagonal combinations. In the simplest renormalizable case, this requires both BEH bosons $H^{\bf\bar 5}~{\rm and}~H'{}^{\bf\bar 5}$ to transform as triplets of $T_7$.} The details are beyond the scope of this paper and will be discussed elsewhere.   


\vskip .5cm
\noindent {\large Seesaw Sector}
\vskip .5cm
\noindent In the TBM texture, the seesaw neutrino mass formula becomes

\begin{align} \label{neutrinomasses}
    M^{}_\nu = Y^{(0)} \frac{1}{\mc{M}} Y^{(0)T} = \mathcal{U}_{\mathrm{TBM}} \: \mc{D}_\nu \: \mc{U}_{\mathrm{TBM}}^T
\end{align}
where $\mc{M}$ is the Majorana mass matrix of the right-handed neutrinos, and $\mc{D}_\nu = \mathrm{diag}(m_1, m_2, m_3)$ is the diagonal light neutrino mass matrix. The numerator $Y^{(0)}$ is the neutral lepton Dirac Yukawa matrix which, in $SO(10)$,  is most simply related to the up-quark Yukawa matrix  $Y^{(0)} \sim Y^{(2/3)}.$

$Y^{(0)}$ inherits the large hierarchy of the up-quark sector\footnote{Whiffs of $\epsilon\approx \lambda^4$ in the seesaw sector are too small to affect seesaw simplicity in generating the reactor angle.}.
This  hierarchy is not replicated by the light neutrino data, and Eq.\eqref{neutrinomasses} implies a correlated squared $\epsilon$ hierarchy in the Majorana matrix.

\noindent We therefore separate out the hierarchy from the Majorana matrix

\be
\mc{M} = Y^{(0)} \mc{M'} Y^{(0)T}.
\ee
By using Eq.(\ref{neutrinomasses}), we can express the Majorana matrix in terms of neutrino masses and the \cancel{CP}-phase,

\be \label{majorana}
    \mc{M'} = \mc{U}_{\mathrm{TBM}}^* \: \mc{D}_\nu^{-1} \: \mc{U}_{\mathrm{TBM}}^\dagger.
\ee
\vskip .5cm
The light neutrino masses are not yet known, although they are bounded by cosmology \cite{planck2015} and  oscillations,
  
\begin{align}
    &m_1 \leq 71.17 \:  ;\quad
    8.68 \: \leq m_2 \leq 71.70 \: ; \quad 
    50.3 \:  \leq m_3 \leq 87.13 ,\: ~({\rm meV})
\end{align}
with
 
\be\label{limits}
58.9\leq m_1^{}+m_2^{}+m_3^{} \leq  230,~({\rm meV}),
\ee
for the normal hierarchy.
 
\vskip .5cm
\noindent Eq.\eqref{majorana} yields
\vskip .5cm

\begin{align*}
 \mc{M'} &= 
\frac{1}{3m_1m_2m_3}\times\\
\scriptstyle
&\begin{pmatrix}
m_3(m_1+2m_2)& m_3(m_1-m_2) & e^{-i \delta }m_3(m_2-m_1)\cr
 m_3(m_1-m_2) & \frac{1}{2}(3m_1m_2
 +2m_1m_3+m_2m_3) & \dfrac{e^{-i \delta }}{2}(3m_1m_2
 -2m_1m_3-m_2m_3) \cr
 e^{-i \delta }m_3(m_2-m_1) & \dfrac{e^{-i \delta }}{2}(3m_1m_2
 -2m_1m_3-m_2m_3) & \dfrac{e^{-2i \delta }}{2}(3m_1m_2
 +2m_1m_3+m_2m_3)
\end{pmatrix}
\end{align*}
It depends on the phase and its sign, although its matrix elements are not yet fixed by experiment. Eq.(\ref{limits}) shows that less than one order of magnitude improvement on the cosmological bound will (hopefully soon) result in an actual measurement.

It is a challenge to theories  to  predict the neutrino masses. For example, all it takes is a Gatto-like relation between the solar angle and $m_1/m_2$ \cite{muterm} to determine that physics.


\section{Conclusion}
This paper has presented a grand-unified asymmetric texture for the Yukawa matrices of the Standard Model. {\color{black}With five free parameters in the input Yukawa matrices, it is designed to reproduce the three CKM angles, the Gatto relation and the GUT scale relations between three down-quark and three charged lepton masses.}

Here neutrino masses are generated by the seesaw mechanism. In the belief that gauge simplicity at GUT scale should be matched by ``seesaw simplicity'' where only large angles appear, we assume  TBM  diagonalization of the seesaw neutrino matrix. Seesaw simplicity  requires the small PMNS reactor angle $\theta_{13}$ to be generated through the charged lepton mixings.

Symmetric electroweak textures fall short of ``seesaw simplicity''. However, in this asymmetric texture the reactor angle $\theta_{13}$ exceeds its pdg value, while the charged lepton mixing contribution to the solar and atmospheric angles yield values outside their pdg allowances.
\vskip .5cm
A \cancel{CP}-phase in the TBM matrix reduces the reactor angle value and drives the solar and atmospheric angles in the right direction. It is noteworthy that it provides one solution for three problems:
\vskip .2cm
\noindent - A \cancel{CP}-phase with $\delta=\pm 78^\circ$ in the TBM matrix lowers $\theta_{13}$ to its experimental value. 
\vskip 0.2cm
\noindent - This corresponds to the PMNS phase $\delta_{CP}=\pm 1.32\pi$ and Jarlskog-Greenberg invariant $J=\mp 0.028$, with magnitude in perfect agreement with experiment \cite{de2018neutrino}.
\vskip .2cm
\noindent - The very same \cancel{CP}-phase adjusts the solar and atmospheric neutrino angles to within one degree of their pdg values.
\vskip .2cm  

{\color{black} Therefore, introducing three input parameters (two nonzero angles and one phase) in the form of a complex TBM matrix enables us to explain four parameters (three mixing angles and one \cancel{CP} phase) in the PMNS matrix.}
The sign of the phase  is a property of the Majorana mass matrix of the right-handed neutrinos, and is not determined by the texture.

We expect that the electroweak side of our texture can be applied to Golden Ratio \cite{goldenratio, *everett2009icosahedral} seesaw diagonalization as well. The next step is to find a common organizing principle that relates the seesaw Majorana matrix to the Standard Model Yukawa matrices. We hope to address this question in a future work.

\section{Acknowledgements}
We thank Gaoli Chen for helpful discussions at the early stage of this work. This research was supported in part by the Department of Energy
under Grant No. DE-SC0010296.

\newpage
\numberwithin{equation}{section}
\appendix
\section*{Appendices}
\section{Symmetric-Antisymmetric Textures}\label{appA}
We first consider textures with only symmetric and/or antisymmetric $\mathbf{5}$ and/or $\mathbf{45}$ couplings. Our objective is to find out textures based on $SU(5)$ grand unification that can reproduce mass relations and mixing angles in down-quark and charged lepton sectors.

\vskip 0.2 cm
\noindent For simplicity, 

\noindent - consider all couplings are real

\noindent - let a single parameter $c'$ denote one diagonal or a pair of off-diagonal $\mathbf{45}$ coupling(s), all other couplings are $\mathbf{5}$, denoted by $a', b', d', g'$ etc. Off-diagonal symmetry/antisymmetry is denoted by sign parameters $\varsigma_{a'}=\pm 1$ etc. All couplings will be expressed in integer powers of the Wolfenstein parameter $\lambda$ with a prefactor: $a' = a \lambda^n$ etc.

\noindent - taking hint from $m_b \approx m_\tau$ at GUT scale, the $(33)$ coupling is assumed to be $\mathbf{5}$ and all other couplings are normalized by this.
\vskip 0.2cm
An important observation is, $\det Y^{(-1/3)}$ should be independent of $c'$ so that it approximates $\det Y^{(-1)}$ at GUT scale.

\vskip 0.2 cm
\noindent Classify these textures as: (i) $\mathbf{45}$ couplings in off-diagonal entries, (ii) $\mathbf{45}$ coupling in  diagonal  entry.

\vskip 0.5cm
\noindent{\bf Off-diagonal $\mathbf{45}$ couplings}
\vskip 0.2cm
\noindent Consider a pair of off-diagonal $\mathbf{45}$ couplings, either symmetric or antisymmetric. There can be three such textures.
\vskip 0.3 cm
\noindent{\large $(12)-(21)$ $\mathbf{45}$ texture}
\begin{align}
    Y^{(-1/3)} &= \left(
\begin{array}{ccc}
 a' & c' & b' \\
 \varsigma_{c'} c' & g' & d' \\
 \varsigma_{b'}b' & \varsigma_{d'} d' & 1 \\
\end{array}
\right)
\end{align}
 $$\det{Y^{(-1/3)}} = -\varsigma_{c'}c'^2-\varsigma_{b'}b'^2 g'+a'(g'-\varsigma_{d'}d'^2)+b'c'd'(\varsigma_{b'}+\varsigma_{c'}\varsigma_{d'}).$$ This can not be made independent of $c'$; thus this texture cannot yield correct mass relations and will not be pursued further.

\vskip 0.3cm
\noindent {\large $(13)-(31)$ $\mathbf{45}$ texture}
\begin{align}
    Y^{(-1/3)} &= \left(
    \begin{array}{ccc}
        a' & b' & c'  \\
        \varsigma_{b'}b' & g' & d'\\
        \varsigma_{c'}c' & \varsigma_{d'}d' & 1
    \end{array}
    \right) 
\end{align}
 $$\det Y^{(-1/3)} = -\varsigma_{b'}b'^2-\varsigma_{c'}c'^2 g'-\varsigma_{d'}a' d'^2+a' g'+b'c'd'(\varsigma_{c'}+\varsigma_{b'}\varsigma_{d'}),$$ which would be independent of $c'$ if $g'=0$ and $(\varsigma_{b'}, \varsigma_{d'}, \varsigma_{c'})$ $=$ $(1,\pm 1, \mp 1)$ or $(-1, \pm 1, \pm 1)$.  With these constraints the texture takes the form
\begin{align}
    Y^{(-1/3)} &= \left(
    \begin{array}{ccc}
        a' & b' & c'  \\
        \varsigma_{b'} b' & 0 & d'\\
        \varsigma_{c'}c' & \varsigma_{d'}d' & 1
    \end{array}
    \right)
\end{align}
Then,
\begin{align} \label{YYdd}
    &Y^{(-1/3)} \, Y^{(-1/3)T} = \mathcal{U}^{(-1/3)} \, \mathcal{D}^{(-1/3)} \, \mathcal{D}^{(-1/3)} \, \mathcal{U}^{(-1/3)T} \nonumber \\
    &= \left(
\begin{array}{ccc}
 a'^2+b'^2+c'^2 & c' d'+\varsigma_{b'}a' b' & \varsigma_{c'}a' c'+c'+\varsigma_{d'}b' d' \\
 c' d'+\varsigma_{b'}a' b' & b'^2+d'^2 & d'+\varsigma_{b'}\varsigma_{c'}b' c' \\
 \varsigma_{c'}a' c'+c'+\varsigma_{d'}b' d' & d'+\varsigma_{b'}\varsigma_{c'}b' c' & c'^2+d'^2+1 \\
\end{array}
\right)
\end{align}
In our chosen basis, $\mathcal{U}^{(-1/3)} = \mathcal{U}_{\,CKM}$.
Then, the Wolfenstein parametrization  of the CKM matrix, ignoring \cancel{CP} phase,
\begin{align} 
\mathcal{U}_{\mathrm{CKM}} &=
    \left(
\begin{array}{ccc}
 1-\dfrac{\lambda ^2}{2} & \lambda  & A \lambda ^3 \sqrt{\eta ^2+\rho ^2} \\
 -\lambda  & 1-\dfrac{\lambda ^2}{2} & A \lambda ^2 \\[0.5em]
 A \lambda ^3 & -A \lambda ^2 & 1 \\
\end{array}
\right) \label{ckm2}
\end{align}
yields, up to leading order,
\begin{align}
    \hspace{-2mm} Y^{(-1/3)} \, Y^{(-1/3)T} &= \mathcal{U}_{\mathrm{CKM}} \, \mathcal{D}^{(-1/3)} \, \mathcal{D}^{(-1/3)} \, \mathcal{U}_{\mathrm{CKM}}^T\\[0.5em]
    &= \left(
\begin{array}{ccc}
 \dfrac{\lambda ^6}{9} &  A^2 \lambda ^5 \sqrt{\eta ^2+\rho ^2} & A \lambda ^3 \sqrt{\eta ^2+\rho ^2} \\[0.8em]
  A^2 \lambda ^5 \sqrt{\eta ^2+\rho ^2} & A^2 \lambda ^4 & A \lambda ^2 \\[0.8em]
 A \lambda ^3 \sqrt{\eta ^2+\rho ^2} & A \lambda ^2 & 1 \\
\end{array}
\right)\label{ydckm}
\end{align}

\noindent Comparing this with Eq.\eqref{YYdd} we observe that $c' = c \lambda^4$, $d' = d \lambda^2$, $a'= a\lambda^3$ and $b' = b\lambda^3$, where $c$, $d$ $\sim$ $\mathcal{O}(1)$ and $a$, $b$ $\lesssim$ $\mathcal{O}(1)$ parameters.
\vskip 0.2cm
The eigenvalues of $ Y^{(-1/3)}Y^{(-1/3)T}$ are the mass-squared of the down-quarks: $m_d^2$, $m_s^2$ and $m_b^2$. These are related by
\begin{align}
    m_d^2 + m_s^2 + m_b^2 &= 1 +2 d^2 \lambda ^4 +  \left(a^2+2 b^2\right)\lambda ^6+ 2 c^2 \lambda ^8 \label{antisym1221sum}\\
    m_d^2\: m_s^2 + m_s^2 \: m_b^2 + m_b^2\: m_d^2 &=  \left(a^2+2 b^2\right)\lambda ^6+d^4 \lambda ^8+ \mathcal{O} \left(\lambda ^{10}\right) \\
    m_d^2\: m_s^2\: m_b^2 &= b^4 \lambda ^{12} + \mathcal{O}\left(\lambda^{13}\right)\label{antisym1221mul}
\end{align}

\noindent Interestingly, Eqs.\eqref{antisym1221sum}-\eqref{antisym1221mul} do not contain any sign ambiguity, therefore, irrespective of sign, we derive
\begin{align}
    m_b^2 &= 1+2d^2 \lambda^4 + (2c^2-d^4)\lambda^8 + \mathcal{O}\left(\lambda^{10}\right)
\end{align}
leaving
\begin{align}
    m_d^2 + m_s^2 &=  \left(a^2+2 b^2\right)\lambda ^6+ \mathcal{O}\left(\lambda^8\right) \label{sum12}
\end{align}

\noindent Eigenvalues of $Y^{(-1)} Y^{(-1)T}$, labelled by $m_e^2$, $m_\mu^2$, $m_\tau^2$, can be derived from those of $Y^{(-1/3)} Y^{(-1/3)T}$ by replacing $c\rightarrow-3c$. This predicts $m_d^2 + m_s^2 = m_e^2 + m_\mu^2$ from Eq.\eqref{sum12}, which is unsatisfactory. Therefore, this texture with off-diagonal $\mathbf{45}$ couplings in $(13)-(31)$ position doesn't yield correct masses for charged leptons and down-quarks. 

\vskip 0.5cm
\noindent {\large $(23)-(32)$ $\mathbf{45}$ texture}
\vskip 0.3 cm
\noindent Proceeding as the previous case, this texture has the following form
\begin{align}
    Y^{(-1/3)} &= \left(
    \begin{array}{ccc}
        0 & b\lambda^3 & d \lambda^4  \\
        \varsigma_{b}b\lambda^3 & g\lambda^2 & c \lambda^2\\
        \varsigma_{d}d \lambda^4 & \varsigma_{c}c \lambda^2 & 1
    \end{array}
    \right) 
\end{align}
subject to the constraints $(\varsigma_{c'}, \varsigma_{b'}, \varsigma_{d'})$ $=$ $(1,\pm 1, \mp 1)$ or $(-1, \pm 1, \pm 1)$. Here $c, d \sim \mathcal{O}(1)$ and $g$, $b$ $\lesssim$ $\mathcal{O}(1)$. Solving the eigenvalues of $Y^{(-1/3)} Y^{(-1/3)T}$ yields
\begin{align}
    m_b^2 = 1+2c^2\lambda^4 + 2 \varsigma_{c'}c^2 g \lambda^6+(2d^2-c^4)\lambda^8 +\mathcal{O}\left(\lambda^9\right)
\end{align}
This leaves
\begin{align}
    m_d^2 + m_s^2 &= g^2 \lambda^4 + (2b^2 -2 \varsigma_{c'}c^2 g)\lambda^6+c^4\lambda^8 \label{sum12tex3}
\end{align}
The dominant term in Eq.\eqref{sum12tex3} is $g^2 \lambda^4$. This suggests that $m_s^2 \approx g^2 \lambda^4 = \lambda^4/9$ at GUT scale, with $g=1/3$. Then, for $Y^{(-1)}$, we will derive $m_\mu^2 \approx  g^2 \lambda^4 = \lambda^4/9$, much smaller than the expected value $\lambda^4$ at GUT scale. This shows that the off-diagonal $\mathbf{45}$ in the $(23)-(32)$ position also fails to generate the correct mass relations. 

\vskip 0.5 cm
\noindent {\bf Diagonal $\mathbf{45}$ coupling}
\vskip 0.2 cm
\noindent Next, we discuss textures with a single $\mathbf{45}$ coupling in in either $(11)$ or $(22)$ position of $Y^{(-1/3)}$. 
\vskip 0.5cm
\noindent {\large (11) $\mathbf{45}$ texture}
\begin{align}
Y^{(-1/3)} =
    \left(
\begin{array}{ccc}
 c \lambda^3 & a \lambda^3 & b \lambda^4 \\
 \varsigma_{a} a \lambda^3 & \varsigma_{g}g^2 \lambda^4 & g \lambda^2 \\
 \varsigma_{b} b \lambda^4 & \varsigma_{g} g \lambda^2 & 1 \\
\end{array}
\right)
\end{align}
where $b, g \sim \mathcal{O}(1)$ and $a$, $c$ $\lesssim$ $\mathcal{O}(1)$.

\noindent Solving the eigenvalue equations of $Y^{(-1/3)}Y^{(-1/3)T}$ gives
\begin{align}
    m_b^2 &= 1+2g^2\lambda^4 + (2 b^2+g^4)\lambda^8 + \mathcal{O}\left(\lambda^9\right)\\
    m_d^2 + m_s^2 &= (2a^2+c^2)\lambda^6 + \mathcal{O}\left(\lambda^8\right) \label{mdms}
\end{align}
irrespective of signs of prefactors. 

\noindent Since $a$, $c$ $\lesssim$ $\mathcal{O}(1)$, Eq.\eqref{mdms} is unable to produce $m_s = \lambda^2/3$ at GUT scale. This will, in turn, predict lower mass of $m_\mu$. Therefore, this texture cannot generate correct masses for down-quarks and leptons. 

\vskip 0.5 cm
\noindent {\large (22) $\mathbf{45}$ texture}
\vskip 0.3 cm
\noindent An analysis parallel to the $(13)-(31) \: \mathbf{45}$ texture results in the following form
\begin{align}
    Y^{(-1/3)} &= \left(
    \begin{array}{ccc}
        0 & a \lambda^3 & b \lambda^3  \\
        \varsigma_{a}a \lambda^3 & c \lambda^2 & g \lambda^2\\
        \varsigma_{b}b \lambda^3 & \varsigma_{g}g \lambda^2 & 1
    \end{array}
    \right) + \mathcal{O} \left(\lambda^6\right) 
\end{align}
where $b$, $g$ $\sim$ $\mathcal{O}(1)$ and $a, c \lesssim \mathcal{O}(1)$. 
\vskip 0.3 cm
\noindent 
Irrespective of the sign of prefactors, this texture produces the same mass relations and mixing angles as the Georgi-Jarlskog texture discussed in section \ref{GJ sym}. It should be noted that the PMNS angles $\theta_{23}$ and $\theta_{12}$ are not too far off from pdg values in this texture.

\vskip 0.5cm
The above discussion of this appendix shows that the texture with $\mathbf{45}$ coupling in the $(22)$ position can, unlike the others, generate the correct mass relations. This implies that the asymmetry must lie in the $\mathbf{5}$ couplings.

\newpage
\section{Asymmetric Textures}\label{appB}
Following appendix \ref{appA}, we discuss how to introduce asymmetry in the $\mathbf{5}$ couplings. Decomposing the charged lepton diagonalizing matrix into rotation matrices
\[\mathcal{U}^{(-1)}=\mathcal{R}_{23}(\phi_{23})\: \mathcal{R}_{13}(\phi_{13}) \: \mathcal{R}_{12}(\phi_{12})\]
we recall that in the symmetric texture, 
\[\phi_{23}=A\lambda^2,\quad \phi_{13}=b\lambda^3,\quad \phi_{12}=-\dfrac{\lambda}{3},\quad \text{where }b=A\sqrt{\rho^2+\eta^2}.\]
Asymmetry can be incorporated by changing these relationships. 
\vskip 0.3cm
\noindent For simplicity, let's change one angle at a time, and inspect how the PMNS matrix is affected.

\vskip .5cm
\noindent {\large - Change $\phi_{23}$}
\vskip .3cm
\noindent In this case we keep $\phi_{13}=b\lambda^3$ and $\phi_{12}=-\dfrac{\lambda}{3}$, while $\phi_{23}$ is unspecified,

\begin{equation}
\begin{aligned}
\mathcal{U}^{(-1)}&=\left(
\begin{array}{ccc}
1 & 0 & 0 \\
0 & c_{23} & s_{23} \\
0 & -s_{23} & c_{23} \\
\end{array}
\right)\left(
\begin{array}{ccc}
1 & 0 & b\lambda^3 \\
0 & 1 & 0 \\
-b\lambda^3 & 0 & 1 \\
\end{array}
\right)\left(
\begin{array}{ccc}
1-\dfrac{\lambda^2}{18} & -\dfrac{\lambda}{3} & 0 \\[0.5em]
\dfrac{\lambda}{3} & 1-\dfrac{\lambda^2}{18} & 0 \\[0.5em]
0 & 0 & 1 \\
\end{array}
\right)\\
&=\left(
\begin{array}{ccc}
1-\dfrac{\lambda ^2}{18} & -\dfrac{\lambda }{3} & b \lambda ^3 \\[0.5em]
\dfrac{c_{23} \lambda }{3}-s_{23}b \lambda ^3 & c_{23}-\dfrac{c_{23} \lambda ^2}{18}+\dfrac{1}{3} s_{23}b \lambda ^4 & s_{23} \\[0.5em]
-\dfrac{s_{23} \lambda }{3}-c_{23}b \lambda ^3 & -s_{23}+\dfrac{s_{23} \lambda ^2}{18}+\dfrac{1}{3} c_{23}b \lambda ^4 & c_{23} \\
\end{array}
\right)+\mathcal{O}\left(\lambda^5\right)
\end{aligned}
\end{equation}
where $c_{23}=\cos \phi_{23}$, $s_{23}=\sin \phi_{23}$.

\noindent Together with TBM seesaw diagonalization, this yields the reactor angle
\[
|\sin \theta_{13}|=\dfrac{1}{\sqrt{2}}\left|\left(\dfrac{c_{23} \lambda }{3}-s_{23}b \lambda ^3\right)+\left(-\dfrac{s_{23} \lambda }{3}-c_{23}b \lambda ^3\right)\right|
\approx
\dfrac{\lambda}{3\sqrt{2}}|c_{23}-s_{23}|\leq \dfrac{\lambda}{3}
\]
which is much smaller than the experimental value $(0.145)$, no matter how we change $\phi_{23}$. 
Therefore modifying $\phi_{23}$ does not help much.

\vskip .5cm
\noindent {\large - Change $\phi_{12}$}
\vskip .5cm
\noindent
\begin{equation}
\label{U12}
\begin{aligned}
\mathcal{U}^{(-1)} = \left(
\begin{array}{ccc}
c_{12} & s_{12} & b\lambda^3 \\
-s_{12} & c_{12} & A\lambda^2 \\
s_{12}A\lambda^2-c_{12}b\lambda^3 & -c_{12}A\lambda^2-s_{12}b\lambda ^3 & 1 \\
\end{array}
\right)+\mathcal{O}\left(\lambda ^5\right)
\end{aligned}
\end{equation}
where $c_{12}=\cos \phi_{12}$, $s_{12}=\sin \phi_{12}$.

\noindent Fitting
\[|\sin \theta_{13}|=\dfrac{1}{\sqrt{2}}\left|-s_{12}+s_{12}A\lambda^2-c_{12}b\lambda^3\right|\]
to the experimental value, we obtain two solutions $s_{12}=0.210\text{ or }-0.217$.

\noindent Plugging $s_{12}$ into Eq.\eqref{U12} and using TBM seesaw diagonalization, the other two angles are now fully determined. For $s_{12}=0.210$,
\begin{equation*}
\theta_{12}=26.41^\circ \: (7.24^\circ \mathrm{\:below\: pdg\: value}),\quad \theta_{23}=42.03^\circ \: (3.55^\circ \mathrm{\:below\: pdg\: value})
\end{equation*}
\vskip 0.2 cm
\noindent For $s_{12}=-0.217$,
\begin{equation*}
\theta_{12}=44.71^\circ \: (11.06^\circ \mathrm{\:above\: pdg\: value}), \quad \theta_{23}=42.03^\circ \: (3.55^\circ \mathrm{\:below\: pdg\: value})
\end{equation*}

\noindent With these large discrepancies in both cases, modifying $\phi_{12}$ does not alleviate the problem. 

\vskip .5cm
\noindent {\large - Change $\phi_{13}$}
\vskip .2cm
\noindent
\begin{equation}
\label{U13}
\begin{aligned}
\mathcal{U}^{(-1)}
&=\left( \!\!\!
\begin{array}{ccc}
c_{13}-\dfrac{c_{13} \lambda ^2}{18} & \dfrac{c_{13} \lambda }{3} & s_{13} \\[0.5em]
-\dfrac{\lambda}{3}-s_{13}A\lambda^2+\dfrac{s_{13}A\lambda^4}{18} & 1-\dfrac{\lambda ^2}{18} -\dfrac{s_{13}A\lambda^3}{3} & c_{13}A \lambda ^2 \\[0.8em]
-s_{13}+\dfrac{s_{13}\lambda ^2}{18}+\dfrac{A\lambda^3}{3} & -\dfrac{s_{13}\lambda }{3} -A\lambda^2+\dfrac{A\lambda^4}{18} & c_{13} \\
\end{array} \!\!\!
\right)+\mathcal{O}\left(\lambda ^5 \right)
\end{aligned}
\end{equation}
where $c_{13}=\cos \phi_{13}$, $s_{13}=\sin \phi_{13}$.

\noindent Following the same procedure, fitting
\[|\sin \theta_{13}|=\dfrac{1}{\sqrt{2}}\left|\left(-\dfrac{\lambda}{3}-As_{13}\lambda^2+\dfrac{As_{13}\lambda^4}{18}\right)+\left(-s_{13}+\dfrac{s_{13}\lambda ^2}{18}+\dfrac{A\lambda^3}{3}\right)\right|\]
to the pdg value yields $s_{13}=-0.267\text{ or }0.128$
\vskip 0.2 cm
\noindent For $s_{13}=-0.267$, both $\theta_{12}$  and $\theta_{23}$  fall short of their experimental value by $12.15^\circ$ and  $1.29^\circ$ respectively.
\vskip 0.1 cm
\noindent For $s_{13}=0.128$, $\theta_{12}$ is $3.44^\circ$ above the experimental value and $\theta_{23}$ is $3.02^\circ$ below its experimental value.

\vskip 0.2cm
\noindent To conclude,
\begin{itemize}
	\item $\theta_{13}$ is too small if we only change $\phi_{23}$. 
	\item If we only change $\phi_{12}$, $\theta_{13}$ can be fitted to its pdg value. But then $\theta_{12}$ is very far away from experiment ($7.24^\circ$ or $11.06^\circ$).
	\item If we only change $\phi_{13}$, $\theta_{13}$ can be fitted to experiment. Choosing $\phi_{13}$ to be in the first quadrant, $\theta_{12}$ and $\theta_{23}$ deviate much less from their pdg values.
\end{itemize}

\noindent None of these seem particularly correct, although the third one looks more promising.

Another way of looking at this phenomenon is to go directly to the Yukawa matrices. There are three generic asymmetries in the Yukawa matrices, and as we will see that changing $\phi_{13}$ is connected to a particular type, in which $Y^{(-1/3)}_{13}\ne Y^{(-1/3)}_{31}$ dominate the asymmetry.

\vskip 0.3 cm
\noindent{\bf Asymmetric Yukawa matrices}\label{changeY}
\vskip 0.2 cm
\noindent For simplicity, consider one asymmetry at a time.

\vskip .3cm
\noindent {\large $(12)-(21)$ asymmetry}
\vskip .2cm
\noindent
Consider
\begin{equation}
Y^{(-1/3)}=\left(
\begin{array}{ccc}
0 & a \lambda ^m & b \lambda ^3 \\
a' \lambda ^n & c \lambda ^2 & g \lambda ^2 \\
b \lambda ^3 & g \lambda ^2 & 1 \\
\end{array}
\right)
\end{equation}
where $a,a',b,c,g\sim \mathcal{O}(1)$.
\vskip 0.2 cm
\noindent Diagonalizing $Y^{(-1/3)}Y^{(-1/3)T}_{_{}}$, we find that the Cabibbo angle is given by $\dfrac{a^{^{}}}{c^{}}\lambda ^{m-2}$, and the mass squared of down-quark is approximately $\left(\dfrac{aa'}{c}\lambda^{m+n-2}\right)^2_{}$. Fitting these to the correct order of $\lambda$ requires 
$m=n=3$.
Now the Yukawa matrix becomes
\begin{equation}
Y^{(-1/3)}=\left(
\begin{array}{ccc}
0 & a \lambda ^3 & b \lambda ^3 \\
a' \lambda ^3 & c \lambda ^2 & g \lambda ^2 \\
b \lambda ^3 & g \lambda ^2 & 1 \\
\end{array}
\right)
\end{equation}
where $a\ne a'$. Comparing the eigenvalues and eigenvectors of $Y^{(-1/3)}Y^{(-1/3)T}$ with $(\mathcal{D}^{(-1/3)})^2$ and $\mathcal{U}_{\mathrm{CKM}}$, respectively, yield
\begin{align}
    a= \dfrac{1}{3} \: ,\:c=\dfrac{1}{3} \: , \: a'=-\dfrac{1}{3} \: \: \mathrm{and} \: \: g=A \:,\: b=A\sqrt{\rho^2+\eta^2}^{}
\end{align} 
\vskip 0.2cm
\noindent The lepton masses are acquired by $c\rightarrow-3c$. 
\vskip 0.2 cm
\noindent The eigenvectors of $Y^{(-1)}Y^{(-1)T}$ generate $\mathcal{U}^{(-1)}$, which, with TBM seesaw matrix, yields
\[|\sin \theta_{13}|=\dfrac{\lambda}{\sqrt{2}}\left|\dfrac{1 }{3}-\dfrac{A}{3} \lambda ^2\right|\approx \dfrac{\lambda}{3\sqrt{2}}\]
again, one third of the experimental value.
\vskip .5cm
\noindent {\large $(23)-(32)$ asymmetry}
\vskip .5cm
\noindent
\begin{equation}
Y^{(-1/3)}=\left(
\begin{array}{ccc}
0 & a \lambda ^3 & b \lambda ^3 \\
a \lambda ^3 & c \lambda ^2 & g \lambda ^m \\
b \lambda ^3 & g' \lambda^n  & 1 \\
\end{array}
\right)
\end{equation}
where $a,b,c,g,g'\sim \mathcal{O}(1)$.
\vskip 0.2cm
\noindent Comparing the eigenvalues and mixing matrix of $Y^{(-1/3)} Y^{(-1/3)T}$ to GUT scale down-quark masses and $\mathcal{U}_{\mathrm{CKM}}$ yields
\begin{gather}
a=\dfrac{1}{3}, \: c= \dfrac{1}{3},\ g=A,\ m=2,\ b=A\sqrt{\rho^2+\eta^2}
\end{gather}
In the charged lepton sector, diagonalizing $Y^{(-1)}Y^{(-1)T}$, we get $\mathcal{U}^{(-1)}$, together with TBM seesaw diagonalization which yields
\[|\sin \theta_{13}|=\dfrac{\lambda}{3\sqrt{2}}\left|1-g'\lambda ^n-3A\lambda ^2\sqrt{\rho^2+\eta^2}\right|\approx \dfrac{\lambda}{3\sqrt{2}},\]
still one third of the experimental value, no matter what values $g'$ and $n$ take.
\vskip .5cm
\noindent {\large $(13)-(31)$ asymmetry}
\vskip .2cm
\noindent
\begin{equation}
Y^{(-1/3)}=\left(
\begin{array}{ccc}
0 & a \lambda ^3 & b \lambda ^m \\
a \lambda ^3 & c \lambda ^2 & g \lambda ^2 \\
b' \lambda^n  & g \lambda ^2 & 1 \\
\end{array}
\right)
\end{equation}

\noindent In this texture $m=3$ is fixed by the (13) angle of CKM. Furthermore, it can be shown that if $n>1$, the reactor angle
\[|\sin \theta_{13}|=\dfrac{1}{\sqrt{2}}\left|\dfrac{\lambda }{3}+b'\lambda^n\right|\approx \dfrac{\lambda}{3\sqrt{2}}\]
is again one third of its pdg value. 
\vskip 0.2 cm
Considering $m=3$ and $n=1$, this texture has been discussed in detail in section \ref{sec3}. 

\vskip 0.3cm
None of these three types of asymmetries yields satisfactory values for the PMNS angles. However there are important differences. 
\vskip 0.2cm
In the first two cases, when the asymmetries are along $(12)-(21)$ or $(23)-(32)$, the reactor angle is much lower than its pdg value. In these cases TBM diagonalization does not agree with experiment, unless we deviate from it by introducing a new parameter.
\vskip 0.2cm
However, when the asymmetry is along $(13)-(31)$, $\theta_{13}$ is {\bf larger} than its experimental value. As showed in section \ref{sec3}, introducing a phase in TBM reduces $\theta_{13}$, while bringing the other two angles even closer to their pdg central values.

\newpage
\section{Numerical Summary}\label{appC}
The asymmetric texture is expressed in a basis where the up-quark Yukawa matrix is diagonal. The two input Yukawa matrices stem from the $\bf 5$ and $\bf 45$ $SU(5)$ couplings, which yield the two electroweak matrices,

\be
Y^{(-1/3)}=
\begin{pmatrix}
b d \lambda ^4 & a \lambda ^3 & b \lambda ^3 \cr
a \lambda ^3 & c \lambda ^2 & g \lambda ^2 \cr
d \lambda  & g \lambda ^2 & 1 \\
\end{pmatrix}
=
\begin{pmatrix}
0.251\lambda^4 & 0.333\lambda^3 & 0.306\lambda^3 \cr
0.333\lambda^3 & 0.333\lambda^2 & 0.811\lambda^2 \cr
0.822\lambda& 0.811\lambda^2 & 1
\end{pmatrix}
\ee
\be
Y^{(-1)}=
\begin{pmatrix}
b d \lambda ^4 & a \lambda ^3 & d \lambda  \cr
a \lambda ^3 & -3c \lambda ^2 & g \lambda ^2 \cr
b \lambda^3  & g \lambda ^2 & 1 \\
\end{pmatrix}
=
\begin{pmatrix}
0.251\lambda^4 & 0.333\lambda^3 & 0.822\lambda \cr
0.333\lambda^3 & -\lambda^2 & 0.811\lambda^2 \cr
0.306\lambda^3& 0.811\lambda^2 & 1
\end{pmatrix}
\ee
where we have used  Eq.(\ref{prefactors}) to express the prefactors in terms of the Wolfenstein parameters. Note  numerical coincidences between prefactors as $A\approx \sqrt{2/3}$. We summarize their numerical outcomes.

\vskip .5cm
\noindent - Masses of charged leptons and down-quarks up to one overall constant:
\vskip .2cm
\be
m_b=1.019,\,\,
m_d=0.849\times10^{-3}=0.994\dfrac{\lambda ^4}{3},\,\, 
m_s=0.016=0.951 \dfrac{\lambda ^2}{3}. 
\nonumber
\ee
\be
m_\tau=1.019,\,\,
m_e=0.259\times10^{-3}=0.912 \dfrac{\lambda ^4}{9},\,\, 
m_\mu=0.052=1.036 \lambda ^2.
\nonumber \ee
\vskip .5cm
\noindent  - CKM and charged lepton mixing matrices:
\vskip .2cm
\begin{align}
\mc{U}^{}_{\mathrm{CKM}}=\mc{U}^{(-1/3)}&=\left(
\begin{array}{ccc}
0.9751 & 0.2215 & 0.0036 \\
-0.2215 & 0.9743 & 0.041 \\
0.0055 & -0.0407 & 0.9992 \\
\end{array}
\right);\\[1em]
\mathcal{U}^{(-1)}&=\left(
\begin{array}{ccc}
0.9814 & 0.0628 & 0.1816 \\
-0.0709 & 0.9967 & 0.0384 \\
-0.1786 & -0.0505 & 0.9826 \\
\end{array}
\right)
\nonumber
\end{align}
\vskip .5cm
\noindent -  CKM angles: 
\vskip .2cm
\be
\mathcal{U}^{(-1/3)}_{12}=0.985\lambda,\,\,
\mathcal{U}^{(-1/3)}_{13}=1.041A \lambda ^3 \sqrt{\eta ^2+\rho ^2},\,\,
\mathcal{U}^{(-1/3)}_{23}=0.998A \lambda ^2.
\ee

\noindent The Gatto relation holds: ${m_d}/{m_s}=1.045\lambda^2$.
\vskip 0.5cm
\noindent - Reactor angle:
\vskip .2cm

\be\nonumber
|\sin \theta_{13}|=\dfrac{1}{\sqrt{2}}|0.0709+0.1786e^{i\delta}|~\longrightarrow~~\delta=\mp 78^\circ
\ee

\vskip 0.5cm
\noindent - PMNS matrix ($\delta=\mp 78^\circ$):
\vskip .2cm
\begin{equation*}\label{numPMNS}
\mathcal{U}_{\mathrm{PMNS}}=\left(
\begin{array}{ccc}
0.8156 \pm 0.0714 i & 0.5463 \mp 0.1010 i & -0.0754\pm 0.1237 i \\
-0.3598\pm .0202 i & 0.6176 \mp 0.0286 i & 0.6977\pm 0.0350 i \\
0.2128\mp 0.3930 i & 0.0135\pm 0.5559 i & 0.1661\mp 0.6808 i \\
\end{array}
\right).
\end{equation*}
\vskip .5cm
\noindent - PMNS angles:
\vskip .2cm
\be\nonumber
\label{numtheta}
\theta_{13}=8.33^\circ,\,
\theta_{12}=34.16^\circ(0.51^\circ~\text{above pdg}),\,
\theta_{23}=44.91^\circ(0.66^\circ~\text{below pdg}).
\ee
\vskip .5cm
\noindent  - PMNS CP-violating phase \& Jarlskog-Greenberg invariant:
\vskip .2cm
$$
\delta_{CP}=\pm 1.32\pi;\quad J=\mp 0.028,~~{\rm near~pdg}.
$$

\newpage

\bibliography{asymmetric}
\bibliographystyle{rsc}

\end{document}